\documentclass[preprint,aps,showpacs,showkeys,amsmath,amssymb,nopreprintnumbers,nofootinbib]{revtex4}
\usepackage{graphicx}
\usepackage{dcolumn}
\usepackage{bm}

\def\bge{\begin{equation}}
\def\ene{\end{equation}}
\def\bg{\begin{eqnarray}}
\def\en{\end{eqnarray}}

\begin{document}


\title{Scalar and vector meson propagation in asymmetric nuclear matter}

\author{Yukihiro Muto}
\author{Koichi Saito}%
 \email{ksaito@ph.noda.tus.ac.jp}
\affiliation{%
Department of Physics, Faculty of Science and Technology,\\
Tokyo University of Science, Noda 278-8510, Japan 
}%


\date{\today}

\begin{abstract}
The propagation of the scalar ($\sigma$ and $\delta$) and vector ($\omega$ and $\rho$) mesons in an iso-asymmetric 
nuclear matter is studied in detail, using the Walecka model.  
We calculate the invariant masses and spectral functions of the mesons, including the effect of meson mixing. 
At finite density, the mixing effect is quite important in the propagation of the scalar and (longitudinal) vector mesons.  In 
the $\sigma$ channel, we find a three-peak structure in the spectral function, caused by 
the mixing effect.  
\end{abstract}

\maketitle

The variation of hadron properties in hot and/or dense nuclear matter has recently drawn a great deal  
attention.  In particular, 
the medium modification of the light vector ($\rho$, $\omega$ and 
$\phi$) meson masses, which may be a probe to study hadron properties in the extreme matter, 
has been investigated by many authors~\cite{qm06}.  

Experimentally, the photoproduction of $\omega$ mesons on 
a Nb target was recently measured at the ELSA tagged photon facility~\cite{elsa}. 
The result shows that, for low momenta less than $500$ MeV, the in-medium 
$\omega$ meson mass at $0.6 \rho_0$ ($\rho_0$ is the saturation density of nuclear matter) 
is about $8$\% smaller than the free mass.  Furthermore, the experiment to measure the invariant mass spectra 
of $e^+e^-$ pairs produced in $12$ GeV proton-induced nuclear (C and Cu) reactions was also performed 
at the KEK proton synchrotron~\cite{naruki}.  On the low mass side of the 
$\omega$ meson peak, a significant enhancement over the known hadronic sources is observed.  It is well 
understood by a model that takes account of the medium modification for the vector meson mass. 
These experimental results may suggest a downward shift of the vector meson mass in a  
nuclear matter (see also Ref.~\cite{ceres}). 


On the other hand, the $\sigma$ meson has been treated 
as a correlated two-pion state in the scalar channel for a 
long time.  However, some people have argued  
for the existence of the $\sigma$ as a genuine resonance~\cite{sigma}.  Anyway, it is well recognized that the $\sigma$ meson 
is vital to produce the mid-range, attractive force between two nucleons in a matter. 
The $\delta$ (or $a_0$) meson also plays an important role in an iso-asymmetric nuclear matter.

At finite nuclear density, the scalar meson can 
couple to the longitudinal (L) mode of the vector meson~\cite{chin}.  Furthermore, when 
charge symmetry is broken, the iso-scalar meson can couple to the iso-vector one.  
In an iso-asymmetric nuclear matter, where the number of protons, $Z$, is different from that of neutrons, $N$,  
the effect of charge symmetry breaking (CSB) is more enhanced than in vacuum, because 
the difference between the in-medium proton, $M_p^*$, and neutron, $M_n^*$, masses is larger than that in vacuum~\cite{will,mori}. 

This work is an extension and elaboration of studies in Ref.~\cite{saito}, 
where the iso-scalar $\sigma$ and $\omega$ mesons moving in an iso-symmetric 
($Z=N$) nuclear matter were considered.  Here we generalize it in an asymmetric ($Z\neq N$) nuclear matter, 
including the iso-vector $\delta$ and $\rho$ mesons. We do not consider the propagation of 
pseudoscalar ($\pi$ and $\eta$) mesons, because we use the relativistic Hartree approximation (RHA) in the 
calculation. Instead, we include the effect of the scalar or vector meson decay into 
these pseudoscalar mesons. 

We use a purely hadronic model, i.e., quantum hadrodynamics (QHD or the Walecka model)~\cite{sw}. 
The Lagrangian density is given by
\begin{equation}
{\cal L} = {\cal L}_{QHD-I} + {\cal L}_{\delta} + {\cal L}_{\rho} + {\cal L}_{\pi \eta} + \delta {\cal L}_{CT} 
, \label{lag-1}
\end{equation}
where ${\cal L}_{QHD-I}$ is the usual, QHD-I Lagrangian density for the $\sigma$-$\omega$-nucleon system~\cite{sw} 
and $\delta {\cal L}_{CT}$ is a counter term for renormalizations. 

The $\delta$ and $\rho$ mesons are, respectively, described by ${\cal L}_{\delta}$ and ${\cal L}_{\rho}$. 
The $\delta$ meson couples to a nucleon through the isovector, scalar interaction: 
${\cal L}_{\delta}^{int} = g_{\delta} \bar{\psi} \tau_z \delta \psi$, 
where $\psi \, (\delta)$ is the field of the nucleon (the neutral member of $\delta$ meson) and 
$g_\delta$ is the $\delta$-nucleon coupling constant. The nucleon mass in matter is thus given as
$M_{\binom{p}{n}}^* = M_{\binom{p}{n}} - g_{\sigma} \sigma \mp g_{\delta} \delta$ 
with $\sigma$ the $\sigma$ mean field and $g_{\sigma}$ the $\sigma$-nucleon coupling constant. 
Here $M_p (= 938.27$ MeV) and $M_n (= 939.57$ MeV) are, respectively, the proton and neutron masses in vacuum.  
The $\rho$ meson has a tensor as well as a vector coupling to a nucleon, and the interaction is given by 
${\cal L}_{\rho}^{int} = -g_{\rho} \bar{\psi} \gamma_{\mu} \rho^{\mu} \tau_z \psi 
 + (f_{\rho}/2M_i) \bar{\psi} \sigma_{\mu \nu} \partial^{\nu} \rho^{\mu} \tau_z \psi$ , 
where $i =$ proton ($p$) or neutron ($n$), $\rho^{\mu}$ is the neutral $\rho$ meson field 
and $g_{\rho}$ ($f_{\rho}$) is the vector (tensor) coupling constant. 

The pseudoscalar, $\pi$ and $\eta$, mesons are described by ${\cal L}_{\pi \eta}$, in which 
their interactions to a nucleon are, respectively, written by the isovector and isoscalar, pseudoscalar couplings. 

Firstly, we have to solve the nuclear ground state within RHA. The total energy density is given by 
${\cal E}_{tot} = {\cal E}_{RHA} + \Delta {\cal E}_{VF}$, 
where ${\cal E}_{RHA}$ is the usual one in RHA~\cite{sw}. 
The second term comes from the vacuum fluctuation correction (caused by the nucleon loop) to the 
$\sigma$ and $\delta$ meson propagators~\cite{mori,saito}, and it is given by (also see around Eq.(\ref{ct-1}))
\begin{eqnarray}
\Delta {\cal E}_{VF} &=& \frac{1}{4 \pi^2} \sum_{i=p,n} \left[ M_i^2 \left\{ \frac{(M_p+M_n)-(M_p^*+M_n^*)}{2} \right\}^2 
  \left( \frac{m_{\sigma}^2}{4 M_i^2} - 3 +\frac{3}{2} f(y_{\sigma(i)}) \right) \right. \nonumber \\
  &+& M_i^2 \left\{ \frac{(M_p-M_n)-(M_p^*-M_n^*)}{2} \right\}^2 
   \left( \frac{m_{\delta}^2}{4 M_i^2} - 3 +\frac{3}{2} f(y_{\delta(i)}) \right) \nonumber \\
 &-&  \left. \frac{1}{4} (M_i-M_i^*)^4 \ln \frac{M_i}{2M-M_i} \right] , \label{vf2} 
\end{eqnarray}
where $m_{\sigma (\delta)}$ is the $\sigma \, (\delta)$ meson mass, 
$y_{j(i)} = 1 - (4 M_i^2/m_j^2)$ with $m_j$ the meson mass ($j = \sigma$ or $\delta$), $f(y) = 2 \sqrt{-y} \tan^{-1}\sqrt{-y^{-1}}$ and 
$M = (M_p+M_n)/2$.  

The effective nucleon mass, $M_i^*$, in matter is self-consistently calculated by solving 
the following, two equations: 
\begin{equation}
\rho_{s(p)} \pm \rho_{s(n)} - \frac{m_{j}^2}{g_{j}^2} \left\{ \frac{(M_p \pm M_n)-(M_p^* \pm M_n^*)}{2} \right\} 
+ \Delta M_{p} \pm \Delta M_{n} = 0 , \ {\rm for} \ j= \binom{\sigma}{\delta} , \label{enmcond}
\end{equation}
where $\rho_{s(i)}$ is the proton or neutron scalar density in matter and 
\begin{eqnarray}
\Delta M_{i} &=& - \frac{1}{2 \pi^2} \left[ M_i^{*3} \ln \left( \frac{M_i^*}{M_i} \right) + M_i^2 ( M_i - M_i^* ) 
- \frac{5}{2} M_i ( M_i - M_i^* )^2 + \frac{11}{6} ( M_i - M_i^* )^3 \right] \nonumber \\
&-& \frac{1}{2 \pi^2} \left[  M_i^2 \left\{ \frac{(M_p+M_n)-(M_p^*+M_n^*)}{2} \right\} \left( \frac{m_{\sigma}^2}{4M_i^2} 
-3 + \frac{3}{2} f(y_{\sigma(i)}) \right) \right. \nonumber \\
&\pm&  M_i^2 \left\{ \frac{(M_p-M_n)-(M_p^*-M_n^*)}{2} \right\} \left( \frac{m_{\delta}^2}{4M_i^2} 
-3 + \frac{3}{2} f(y_{\delta(i)}) \right) \nonumber \\
&-&  \left. \frac{1}{2}    (M_i-M_i^*)^3 \ln \frac{M_i}{2M-M_i} \right] , \ \ \  
\rm{for} \ \binom{\rm{proton}}{\rm{neutron}} .
\label{enmcond2}
\end{eqnarray}

A complete investigation of the propagation of $\sigma$, $\omega$, $\delta$ and $\rho$ 
mesons requires the inclusion of the meson mixing effect in a medium.  
To compute it, we sum over the ring diagrams~\cite{chin,will,mori,saito}, which consist 
of repeated insertions of the lowest order, one-loop proper polarization part. It is convenient to use 
the full meson propagator, ${\cal P}_{ab}$ ($a, b = 1 \sim 10$), in the form of a $10 \times 10$ matrix. 
The lowest order meson propagator, ${\cal P}^0$, is then given by a block-diagonal form: 
${\cal P}^0 = \mbox{block-diag} \left( s_0, w_0^{\mu \nu},  d_0,  r_0^{\mu \nu} \right)$, 
where the free propagators for the $\sigma$, $\omega$, $\delta$ and $\rho$ are, respectively, expressed by 
$s_0(q)=p_\sigma(q)$, $w_0^{\mu \nu}(q)=\xi^{\mu \nu}p_\omega(q)$, $d_0(q)=p_\delta(q)$ and 
$r_0^{\mu \nu}(q)=\xi^{\mu \nu}p_\rho(q)$ with 
$\xi^{\mu \nu} = -g^{\mu \nu} + q^{\mu} q^{\nu}/q_{\lambda}^2$ and $p_j(q) = (q_{\mu}^2 - m_j^2 + i\varepsilon)^{-1}$ 
($j=\sigma$, $\omega$, $\delta$ or $\rho$, and $q^\mu$ is the four-momentum).  

Dyson's equation for ${\cal P}$ is also given in matrix form as 
\bge
{\cal P} = {\cal P}^0 + {\cal P}^0 \Pi {\cal P},  \label{full}
\ene
and the polarization insertion, $\Pi$, is written by a $10 \times 10$ matrix 
\bge
\Pi = \left( 
\begin{array}{cccc}
\Pi_{\sigma \sigma}(q) & \Pi_{\sigma \omega}^\nu(q) & \Pi_{\sigma \delta}(q) & \Pi_{\sigma \rho}^\nu(q) \\
\Pi_{\omega \sigma}^\mu(q) & \Pi_{\omega \omega}^{\mu \nu}(q) 
       & \Pi_{\omega \delta}^\mu(q) & \Pi_{\omega \rho}^{\mu \nu}(q) \\
\Pi_{\delta \sigma}(q) & \Pi_{\delta \omega}^\nu(q) & \Pi_{\delta \delta}(q) & \Pi_{\delta \rho}^\nu(q) \\
\Pi_{\rho \sigma}^\mu(q) & \Pi_{\rho \omega}^{\mu \nu}(q) 
       & \Pi_{\rho \delta}^\mu(q) & \Pi_{\rho \rho}^{\mu \nu}(q) 
\end{array} 
\right),
\label{pol}
\ene
where $\Pi_{a b}$ ($a, b = \sigma, \omega, \delta, \rho$) stands for the polarization for the process 
where the $b$ meson is converted into the $a$ meson 
through the nucleon loop.  (Hereafter, $\Pi_{a a}$ is simply denoted by $\Pi_{a}$.)  
The polarization insertion, $\Pi$, can be separated into two subspaces, i.e., the iso-scalar and iso-vector 
block matrices.  The coupling between the two blocks vanishes when charge symmetry is exact. 
However, even in vacuum, there exists the explicit CSB, i.e., the small difference between $M_p$ and $M_n$, and 
the CSB effect is magnified in an asymmetric matter~\cite{mori}. 

Because the nucleon propagator, $G(k)$, is divided into the Feynman (F) and density-dependent (D) pieces, 
the polarization due to the nucleon loop consists of the contribution depending on $\rho_B$ (the nuclear density) 
and that involving only the Feynman piece.  The former (the D part) can be calculated analytically, 
while the latter (the F part) is divergent.  We thus treat the F part using the method of dimensional regularization. 
For the detail of the D part, see Refs.~\cite{lim,mori,saito}. 

To regularize the nucleon loop contributions to the scalar meson propagators and the $\sigma$-$\delta$ mixing, we choose 
the counter term in Eq.(\ref{lag-1}) as
\begin{equation}
\delta {\cal L}_{CT} = \frac{1}{2} Z_{\sigma} \partial_{\mu} \sigma \, \partial^{\mu} \sigma 
+ \frac{1}{2} Z_{\delta} \partial_{\mu} \delta \, \partial^{\mu} \delta
+ \sum_{l=2}^{4} \frac{A_l}{l!} \sigma^l + \sum_{l=2}^{4} \frac{B_l}{l!} \delta^l 
+ \sum_{l,m=1}^{2} \frac{C_{lm}}{l!m!} \sigma^l \delta^m . \label{ct-1}
\end{equation}
To fix the coefficients ($Z_j$, $A_l$, $B_l$ and $C_{lm}$), 
we adopt the following renormalization conditions: 
\begin{eqnarray}
\left. \Pi_j^F \right| _{q_{\mu}^2 = m_j^2 , \sigma = \delta =0} &=& 0  ,  \ \ \ 
\left. \frac{\partial \Pi_j^F}{\partial q_{\mu}^2} \right| _{q_{\mu}^2 = m_j^2 , \sigma = \delta =0} = 0 , \ \ \ 
\left. \frac{\partial \Pi_j^F}{\partial \sigma} \right| _{q_{\mu}^2 = 0 , \sigma = \delta =0} = 0 ,  \nonumber \\
\left. \frac{\partial^2 \Pi_\sigma^F}{\partial \sigma^2} \right| _{q_{\mu}^2 = 0 , \sigma = \delta =0} &=& 0 , \ \ \ 
\left. \frac{\partial \Pi_j^F}{\partial \delta} \right| _{q_{\mu}^2 = 0 , \sigma = \delta =0} = 0 , \ \ \ 
\left. \frac{\partial^2 \Pi_\delta^F}{\partial \delta^2} \right| _{q_{\mu}^2 = 0 , \sigma = \delta =0} = 0 , 
\label{rencond-1} \\
\left. \Pi_{\sigma \delta}^F \right| _{q_{\mu}^2 = 0 , \sigma = \delta =0} &=& 0 , \ \ \ 
\left. \frac{\partial^2 \Pi_{\sigma \delta}^F} {\partial \sigma \partial \delta} \right| _{q_{\mu}^2 = 0 , 
\sigma = \delta =0} 
= 0 , \nonumber
\end{eqnarray}
where $\Pi^F$ is the F part of the polarization insertion. This counter term yields 
the vacuum fluctuation correction, Eq.(\ref{vf2}), to the energy density.  

In addition to the nucleon loop contribution, the lowest order polarization insertion for the $\sigma$ involves 
the pion loop contribution~\cite{saito}
\begin{equation}
\Pi_{\sigma,\pi}(q)= \frac{3}{2} ig_{\sigma\pi}^2 m_\pi^2 \int 
     \frac{d^4k}{(2\pi)^4} \Delta_\pi(k) \Delta_\pi(k+q), \label{pis} 
\end{equation}
with $g_{\sigma \pi}$ the $\sigma$-$\pi$ coupling constant, $m_\pi$ the pion mass and $\Delta_\pi(k)$ the pion propagator. 
The imaginary part of $\Pi_{\sigma,\pi}$ describes the $\sigma$ meson decay into $2\pi$. In contrast, 
the $\delta$ polarization insertion includes the $\eta$-$\pi$ loop contribution~\cite{etapi} 
\begin{equation}
\Pi_{\delta,\pi\eta}(q) = ig_{\delta\pi\eta}^2 \left( \frac{m_\delta^2-m_\eta^2}{m_\pi^2} \right)^2 \int 
     \frac{d^4k}{(2\pi)^4} \Delta_\eta(k) \Delta_\pi(k+q), \label{pidelta} 
\end{equation}
with $g_{\delta\pi\eta}$ the $\delta$-$\pi$-$\eta$ coupling constant and $\Delta_\eta(k)$ the $\eta$ meson propagator 
($m_\eta$ is its mass).  To remove the divergences appearing in these polarizations, we use the 
renormalization conditions on the real parts of the polarizations: 
\begin{equation}
\left. \Re e \, \Pi_{\sigma,\pi}\right|_{q_{\mu}^2 = m_\sigma^2} = 0 , \ \ \ 
\left. \Re e \, \Pi_{\delta,\pi\eta}\right|_{q_{\mu}^2 = m_\delta^2} 
= 0 . \label{smesonrenorm} 
\end{equation}

For the vector meson, the F part of the polarization insertion arising from the nucleon loop, 
$\xi^{\mu \nu} \Pi_{j=\omega, \rho}^{F}$, again contains the divergent piece, and it may be removed by the condition~\cite{saito,shiomi}
\begin{equation}
\left. \Pi_\omega^F \right|_{q_\mu^2 = m_\omega^2, \sigma=\delta=0}  =0 , \ \ \ 
\left. \frac{\partial^n \Pi_\rho^F}{\partial (q_{\mu}^2)^n}\right|_{q_\mu^2 = m_\rho^2, \sigma=\delta=0}  =0 , 
\label{vmesonrenorm}
\end{equation}
where $n = 0, 1, 2, \cdots$.  
The polarization insertion for the $\rho$ meson also involves the pion loop contribution~\cite{urban}
\begin{equation}
\Pi_{\rho,\pi}^{\mu \nu}(q) = i g_{\rho \pi}^2 
\int \frac{d^4 k}{(2 \pi)^4} \left[ (2k+q)^{\mu} (2k+q)^{\nu} \Delta_{\pi}(k+q) \Delta_{\pi}(k) 
 - 2 g^{\mu \nu}  \Delta_{\pi}(k) \right] , 
\label{pirho}
\end{equation}
where $g_{\rho\pi}$ is the $\rho$-$\pi$ coupling constant. This again involves the divergence, but 
the same condition as in Eq.(\ref{smesonrenorm}) may remove it. 

\begin{figure}
\includegraphics[scale=1]{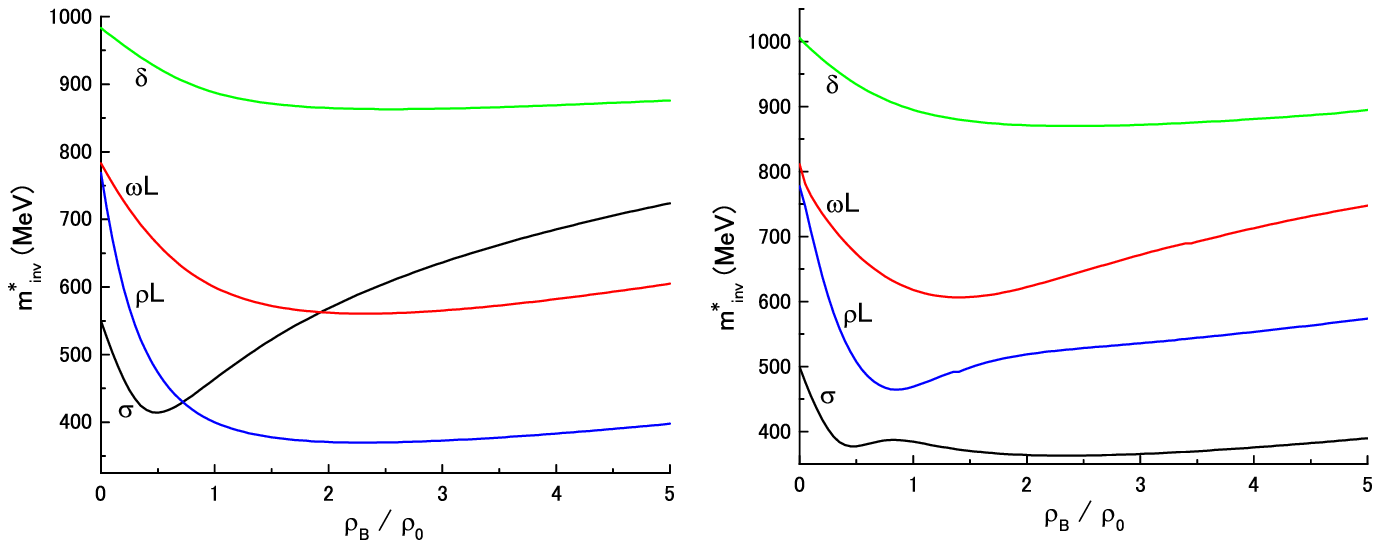}\\
\includegraphics[scale=1]{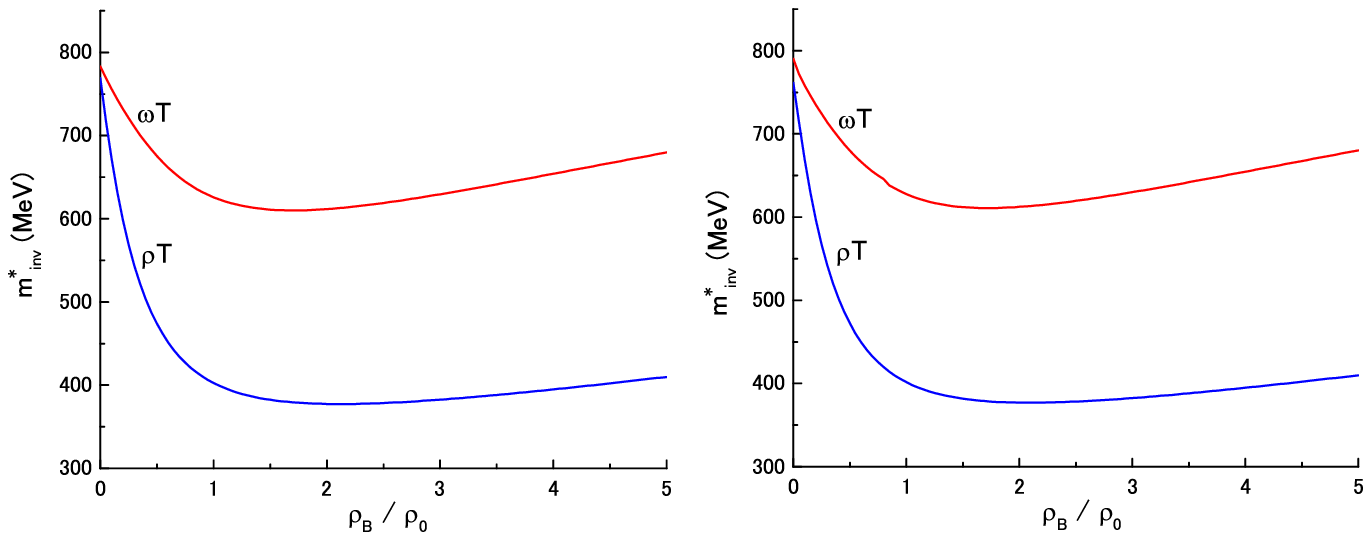}
\caption{\label{invms} Invariant mass with or without the mixing effect ($f_p=0.3$ and $q = 500$ MeV). 
The left, two figures are for the meson masses without 
the mixing effect, while the right ones are for the masses including it. 
}
\end{figure}

For the $\rho$-$\omega$ mixing in vacuum, the polarization insertion due to the 
proton or neutron loop diverges.  A finite mixing amplitude is then obtained by taking the difference between 
proton and neutron contributions~\cite{will}. 
In contrast, the divergent piece arising from the tensor coupling is proportional to $M^*_i/M_i$ in matter, and hence 
it cannot be removed by the same method as in vacuum.  However, 
we simply ignore this divergent piece in the present calculation.\footnote{
The tensor interaction is not a renormalizable one in the conventional sense, and the strict renormalizability 
is not necessary in effective field theories. 
The present prescription is equivalent to taking only the minimal, finite part of $\Pi^{F(t)\mu\nu}_{\rho\omega}$. 
This ambiguity may not affect much the final results of the invariant mass and the spectral function (see 
Figs.~\ref{invms}, 2), because the $\omega$ branch is not close to the $\rho$ branch in the dispersion relation 
and hence the effect of $\rho$-$\omega$ mixing is small in matter. 
} 
In the F part of the polarization insertion, the finite mixing amplitude from the tensor coupling is thus given by 
$\Pi^{F(t)\mu\nu}_{\rho\omega}=\xi^{\mu\nu} (g_\omega f_\rho/8\pi^2) ( K_p - K_n )$, where 
\bge
K_i = \frac{M_i^*}{M_i} q_\mu^2 \int_0^1 dx \, x(1-x) \ln \left[ M_i^{*2} - q_\mu^2 x(1-x) \right] .  \label{K}
\ene

The mixing between the scalar and vector mesons, e.g. $\Pi_{\sigma \omega}$, etc., does not occur 
in vacuum and hence the regularization for such mixing is not necessary~\cite{mori,saito}. 

The dielectric function, $\varepsilon$, in a medium is now defined by~\cite{chin,saito}
\bge
\varepsilon = \mbox{det}(I - {\cal P}^0 \Pi) = \varepsilon_T^2 \times \varepsilon_{SL}, \label{diel}
\ene
where $\varepsilon_{T[SL]}$ is the dielectric function for the 
transverse (T) [scalar (S) and longitudinal (L)] mode.  We find 
\begin{eqnarray}
\varepsilon_T &=& ( 1 - p_\rho \Pi_{\rho}^T ) ( 1 - p_\omega \Pi_{\omega}^T ) 
- p_\rho p_\omega ( \Pi_{\rho \omega}^T )^2 , \label{epstrans}\\
\varepsilon_{SL} &=& \left[ ( 1 - p_\sigma \Pi_{\sigma} ) ( 1 - p_\delta \Pi_{\delta} ) 
- p_\sigma p_\delta ( \Pi_{\sigma \delta} )^2 \right]
  \left[ ( 1 - p_\rho \Pi_{\rho}^L ) ( 1 - p_\omega \Pi_{\omega}^L ) - p_\rho p_\omega ( \Pi_{\rho \omega}^L )^2 
\right] \nonumber \\
 & +& \zeta^2  
\left[ ( 1 - p_\sigma \Pi_{\sigma} ) ( 1 - p_\omega \Pi_{\omega}^L ) p_\delta p_\rho ( \Pi_{\delta \rho}^0 )^2
  + ( 1 - p_\sigma \Pi_{\sigma} ) ( 1 - p_\rho \Pi_{\rho}^L ) p_\delta p_\omega ( \Pi_{\delta \omega}^0 )^2 \right. 
\nonumber \\
 & +& ( 1 - p_\delta \Pi_{\delta} ) ( 1 - p_\omega \Pi_{\omega}^L ) p_\sigma p_\rho ( \Pi_{\sigma \rho}^0 )^2
  + ( 1 - p_\delta \Pi_{\delta} ) ( 1 - p_\rho \Pi_{\rho}^L ) p_\sigma p_\omega ( \Pi_{\sigma \omega}^0 )^2 \nonumber \\
 & +& 2 p_\rho p_\omega \Pi_{\rho \omega}^L \Bigl\{ ( 1 - p_\sigma \Pi_{\sigma} ) p_\delta \Pi_{\delta \omega}^0 
\Pi_{\delta \rho}^0
  + ( 1 - p_\delta \Pi_{\delta} ) p_\sigma \Pi_{\sigma \omega}^0 \Pi_{\sigma \rho}^0 \Bigr\} \nonumber \\
 & +& 2 p_\sigma p_\delta \Pi_{\sigma \delta} \Bigl\{ ( 1 - p_\rho \Pi_{\rho}^L ) p_\omega \Pi_{\sigma \omega}^0 
\Pi_{\delta \omega}^0 
 + ( 1 - p_\omega \Pi_{\omega}^L ) p_\rho \Pi_{\sigma \rho}^0 \Pi_{\delta \rho}^0 \Bigr\} \nonumber \\
 & +& \left. 2 p_\sigma p_\delta p_\rho p_\omega \Pi_{\sigma \delta} \Pi_{\rho \omega}^L ( \Pi_{\sigma \omega}^0 
\Pi_{\delta \rho}^0 
 + \Pi_{\sigma \rho}^0 \Pi_{\delta \omega}^0 )
 + \zeta^2 p_\sigma p_\omega p_\delta p_\rho ( \Pi_{\sigma \omega}^0 \Pi_{\delta \rho}^0
  - \Pi_{\sigma \rho}^0 \Pi_{\delta \omega}^0 )^2 \right] , \label{epslongs}
\end{eqnarray}
with $\zeta^2 = - q_\mu^2 / |{\vec q}\,|^2$,  
$\Pi^{T}( = \Pi^{11} = \Pi^{22})$ the transverse polarization (we choose the direction of ${\vec q}$ as the $z$-axis) and 
$\Pi^{L} (= \Pi^{33}- \Pi^{00})$ the longitudinal one. The free propagator of the $\omega$ meson is now assumed to be 
$p_\omega^{-1}= q_\mu^2 - m_\omega^2 + im_\omega\Gamma_\omega^0$, where we add the width of the $\omega$ 
in vacuum ($\Gamma_\omega^0=9.8$ MeV)~\cite{saito}.  
Using the dielectric function, one can calculate the full propagator, ${\cal P}$ (see Eq.(\ref{full})). 

In the present calculation, we use the following values of the free meson masses: 
$m_\delta=983$ MeV, $m_\sigma=550$ MeV, $m_\omega=783$ MeV, $m_\rho=769$ MeV, $m_\eta=548$ MeV and $m_\pi=138$ MeV. 
So as to fit the saturation condition of normal nuclear matter (${\cal E}_{tot}/\rho_B - M = - 16$ MeV at 
$\rho_0=0.17$ fm$^{-3}$), the symmetry energy ($a_4 = 30$ MeV) and the observed $\rho$-$\omega$ mixing 
amplitude in vacuum ($\langle \rho| H_{mix}|\omega \rangle |_{q_\mu^2 = m_\omega^2} = -4520$ MeV$^2$)~\cite{rm}, we 
determine the coupling constants, $g_\sigma$, $g_\omega$, $g_\rho$ and $g_\delta$. The ratio of the tensor to 
vector coupling constants for the $\rho$ meson is assumed to be $f_\rho/g_\rho = 3.7$, which is suggested by the vector 
dominance model. The coupling constants, $g_{\delta \pi \eta}$ and $g_{\rho \pi}$, are, respectively,  
fixed so as to reproduce the decay widths of the $\delta$ and $\rho$ mesons in vacuum, 
$\Gamma_\delta^0 = 59$ MeV and $\Gamma_\rho^0 = 150$ MeV. Furthermore, we assume the decay width of the $\sigma$ meson in vacuum, 
$\Gamma_\sigma^0 = 500$ MeV, and determine the coupling constant, $g_{\sigma \pi}$, so as to yield it. 
We then obtain the following values of the coupling constants: 
$g_\sigma = 8.119$, $g_\omega = 9.506$, $g_\rho = 4.445$, $g_\delta = 3.732$, 
$g_{\sigma \pi} = 23.65$, $g_{\delta \pi \eta} = 0.4385$ and $g_{\rho \pi} = 6.014$. 
The present calculation gives the (proton-neutron averaged) effective nucleon mass, $M^*/M=0.72$, at $\rho_0$ 
and the incompressibility, $K=465$ MeV.  

Now we are in a position to show our results.  
The condition for determining the collective excitation spectrum is 
equivalent to searching for the zeros of the dielectric function. 
Because we are interested in the medium modification of the meson propagation, we restrict ourselves to 
the meson branch in the time-like region.  

In Fig.~\ref{invms}, we present the ``invariant mass'' as a function of $\rho_B$. The mass is defined by 
$m_{inv}^* \equiv \sqrt{q_0^2 - {\vec q\,}^2}$, where $q_0$ and ${\vec q}$ satisfy the dispersion relation for 
the meson branch. 
In the calculation, the proton fraction, $f_p = Z/(N+Z)$, is chosen to be $0.3$, and the three-momentum 
transfer, $q (= |{\vec q}\,|)$, is $500$ MeV. 
In the figure, we can clearly see the role of mixing. It 
is quite important in determining the meson mass at finite density.  If the mixing is ignored, the masses of 
the L and $\sigma$ modes cross each other and the L mass of the $\rho$ meson is below the other 
ones at high density.  However, when the mixing is included, the L and $\sigma$ masses {\it never} cross 
each other.  At high density, the L mass is pushed upwards, while the 
$\sigma$ mass is pulled downwards.  We can 
understand this phenomenon as a {\it level-level repulsion} due to the 
mixing, which was first studied by Saito {\it et al.}~\cite{saito}.  It is familiar in conventional nuclear physics, e.g. 
in the Nilsson diagram.  For the T mode, 
because the difference between the masses of the $\omega$ and $\rho$ mesons is large, 
the mixing effect is small even at high density. The similar result is obtained 
even in case of $f_p = 0.1$.  

It is very interesting to calculate the spectral functions.  
Because, in a thermal model, the dilepton yields in heavy ion collisions are proportional to the spectral 
functions, the dilepton production rate  per 
unit of four-momentum ($d^4q$) is given by  
$dN_{l^+ l^-}/d^4x d^4q = L_{\mu \nu} H^{\mu \nu}$, where $d^4x$ is the space-time element, and $L_{\mu \nu}$ and 
$H^{\mu \nu}$ are respectively the lepton and the 
hadronic tensors in matter.  The latter is then given in 
terms of the spectral function.  
The spectral function for the $i$-th mode ($i = \sigma, \delta, \omega L, \omega T, \rho L, \rho T$) is 
usually defined by $S_i(m_{inv}^*,q,\rho_B) = - {\Im}m\left[ {\cal P}_i \right]/\pi$. 

%
%
In Fig.~2, we present the spectral function $S_i$ (at $\rho_B/\rho_0=2$ and $f_p=0.3$) 
as a function of $m_{inv}^*$ and $q$. 
The shape for the L or $\sigma$ mode is very complicated. 
In particular, the $\sigma$ mode is quite 
remarkable: at $q$=0 MeV there is only one peak, while at 
finite $q$ {\it three peaks\/} appear in the spectral function.  The two peaks, which have 
relatively large amplitudes, are seen at $m_{inv}^* \simeq m_\sigma$, and 
this two-peak structure is due to the $\sigma$-$\omega$ mixing~\cite{saito}. 
In addition to it, a new, third peak appears around $m_{inv}^* \simeq 0.4$ GeV, which is caused by 
the mixing effect between the iso-scalar and iso-vector mesons. 

In the $\omega$L mode, the large peak is seen above $m_{inv}^* \simeq 0.5$ GeV, that approaches the usual 
$\omega$-meson branch in the limit $\rho_B \to 0$.  
At finite density, the second, small peak appears below $m_{inv}^* \simeq 0.4$ GeV. It is again generated by the 
mixing effect between the iso-scalar and iso-vector mesons, because such structure cannot be seen at 
$f_p = 0.5$~\cite{saito}.  For the $\rho$L mode, in addition to the main peak around $m_{inv}^* \simeq 0.4$ GeV, 
the second, small peak also appears above $m_{inv}^* \simeq 0.5$ GeV.  In case of $f_p=0.1$, 
these second peaks become more clear. 

The spectral function is very simple for the T mode, because the mixing effect is small. 
However, even for the $\omega$T ($\rho$T) mode, the second, 
very small peak appears around $m_{inv}^* \simeq 0.4$ GeV (above $m_{inv}^* \simeq 0.6$ GeV), which cannot be seen in a 
symmetric matter~\cite{saito}.  Thus, it is again due to the effect of iso-scalar and iso-vector mixing.  
In the $\delta$ mode, the spectral function is very simple, because 
the mixing effect on it is quite small.  

The present result is not sensitive to the decay width of the $\sigma$ meson in vacuum. 
We have checked it using $\Gamma_\sigma^0 = 600$ MeV.  
In contrast, the spectral functions strongly depend on $f_p$. In the $\sigma$ mode, 
one can clearly see that the third peak around $m_{inv}^* \simeq 0.4$ GeV grows very quickly as $f_p$ decreases. 

In summary, using QHD, we have studied the propagation of $\sigma$, $\delta$, $\omega$ and $\rho$ mesons 
in an iso-asymmetric nuclear matter.  
We have illustrated that the effect of the meson mixing is quite important in 
the propagation of the $\sigma$, longitudinal $\omega$ and $\rho$ mesons at finite density.  In particular, in 
the $\sigma$ channel, we have found a {\it three-peak} structure in the spectral function at finite three-momentum 
transfer. It is very interesting to measure such novel structure in future experiments.

%
%

%


\end{document}